\documentclass[%
reprint,
amsmath,amssymb,
aps,
]{revtex4-2}
\usepackage{amsmath,amsfonts,amsthm,amssymb}
\usepackage{graphicx}
\usepackage{mathrsfs}
\usepackage{makecell}
\usepackage{setspace}
\usepackage{amsmath}
\usepackage{footnote}
\usepackage{amsfonts}
\usepackage{mathptmx}
\usepackage{makecell}
\usepackage{multirow}
\usepackage{caption}
\usepackage{graphicx,color,subfigure, color}
\usepackage{dcolumn}
\usepackage{bm}
\usepackage{tcolorbox}

\newcommand{\vect}{\mathbf}
\newcommand{\brac}[1]{\left( {#1} \right)}

\begin{document}

\preprint{APS/123-QED}

\title{A statistical theory for polymer elasticity: from molecular kinematics to continuum behavior }

\author{Lin Zhan}
\author{Siyu Wang}
\affiliation{College of Mechanics and Construction Engineering, Jinan University, Guangzhou, 510632, China}

\author{Rui Xiao}
\email{Corresponding author: rxiao@zju.edu.cn}
\author{Shaoxing Qu}
\affiliation{Department of Engineering Mechanics, Zhejiang University, Hangzhou, 310027, China}

\author{Paul Steinmann}
\address{Institute of Applied Mechanics, Friedrich-Alexander-Universit$\ddot{a}$t Erlangen-N$\ddot{u}$rnberg, Egerland Str. 5, 91058 Erlangen, Germany}

\date{\today}

\begin{abstract}
Predicting the macroscopic mechanical behavior of polymeric materials from the micro-structural features has remained a challenge for decades. Existing theoretical models often fail to accurately capture the experimental data, due to non-physical assumptions that link the molecule kinematics with the macroscopic deformation. In this work, we construct a novel Hamiltonian for chain segments enabling a unified statistical description of both individual macromolecular chains and continuum polymer networks. The chain kinematics, including the stretch and orientation properties, are retrieved by the thermodynamic observables without phenomenological assumptions. The theory shows that the chain stretch is specified by a simple relation via its current spatial direction and the continuum Eulerian logarithmic strain, while the probability of a chain in this spatial direction is governed by the new Hamiltonian of a single segment. The model shows a significantly improved prediction on the hyperelastic response of elastomers, relying on minimal, physically-grounded parameters.
\end{abstract}

\maketitle

The high elasticity of soft polymers is widely attributed to their diverse spatial configurations of macromolecules. This understanding has inspired the developments of various single-chain models for polymeric materials \cite{james1943theory,ortiz1999entropic,heussinger2007statistical,di2019direct,buche2020statistical}, many of which rely on predefined chain kinematics, such as the chain end-to-end length or vector.  However, bridging molecular models with continuum-scale behavior poses a critical challenge: how can chain kinematics be directly connected to macroscopic thermodynamic quantities? Addressing this question, particularly the characterization of chain orientation and extension during deformation, is central to advancing the theoretical modeling of polymer hyperelasticity.

Some models employ discretely oriented chains to represent the chain distribution, deriving chain extensions through geometric relations \cite{paul1943statistical,wang1952statistical,arruda1993three,dal2020extended,grasinger2023polymer}. Despite their popularity, discrete-chain models are not able to capture the evolving orientation of chains during deformation. Meanwhile, these models fail to accurately describe the complex mechanical responses under multi-axial loading \cite{dal2021performance,anssari2023large}. Alternatively, the full network model assumes an isotropic distribution of chains, where each direction has its unique kinematics, aligning more closely with the physical picture of isotropic polymers. Except for several non-affine approaches that define the chain's kinematic by specifying the chain force through complex iterative procedures \cite{miehe2004micro,tkachuk2012maximal,verron2017equal,amores2021network,zhan2023new}, the most widely adopted affine approach assumes that a chain deforms in the same manner as a continuum line element \cite{james1943theory,treloar1975physics,wu1993improved,doi2013soft,vernerey2018statistical,buche2020statistical}. 

The affine approach simultaneously describes both chain stretch and orientation using the relation $\vect r=\vect F\cdot\vect r_0$, where $\vect F$ is the deformation gradient tensor, and $\vect r_0$ and $\vect r$ are the chain end-to-end vectors in the undeformed state and the deformed state, respectively \cite{truesdell2004non}. However, this assumption fails to describe the multi-axial stress response observed in experiments \cite{arruda1993three,boyce2000constitutive,dal2021performance,zhan2023new}. Moreover, it appears that the orientation behavior of the chains may not be properly described by the affine assumption, since different soft materials subject to the same deformation have distinct anisotropic chain orientation features. The underlying mechanism arises from the inherent nature of soft materials: molecules are not fixed in the spatial positions due to strong thermal motion. Under external forces, chain segments aim to align with the force direction to minimize potential energy, while thermal motion drives them to orient isotropically to maximize the entropy. These competing mechanisms result in chain orientations that fall between perfect alignment and complete isotropy. Consequently, it is more appropriate to consider the probability of a chain appearing in a specified spatial position $\vect r$ rather than focusing solely on how a specific chain transitions from its material configuration $\vect r_0$ to the spatial configuration $\vect r$.

We start our analysis by considering a freely jointed chain with $N$ segments, where $N$ is also referred as the chain length. Each segment is treated as a rigid rod-like particle with length of $b$. Let $\vect b_i (i=1, 2, ..., N)$ be the configuration vector of the $i$-th segment. The end-to-end vector of the chain is expressed as $\vect r=\sum_{i=1}^{N}\vect b_i$.  If the chain is free of external forces, ${\vect b_i}$ are isotropically oriented and hence $\left<\vect r\right>=N\left<\vect b_i\right>=\vect 0$. \textcolor{black}{Here and henceforth,  $\left<\bullet\right>$ represents the average of the variable $\bullet$.} The mean square of $\vect r$ is given as $\left<\vect r^2\right>=\sum_{i=1}^{N}\sum_{j=1}^{N}\left<\vect b_i\cdot\vect b_j\right>=Nb^2$, where $\left<\vect b_i\cdot\vect b_j\right>=\delta_{ij}b^2$ as $\vect b_i$ and $\vect b_j$ are independent \cite{doi2013soft}. It is therefore convenient to introduce an initial end-to-end distance $r_0=\sqrt Nb$ in the force free state. 

\begin{figure}[h]

	\subfigure[] 
{
	\begin{minipage}{7.5cm}
		\centering       
		\includegraphics[width=75.00mm,height=35.00mm]{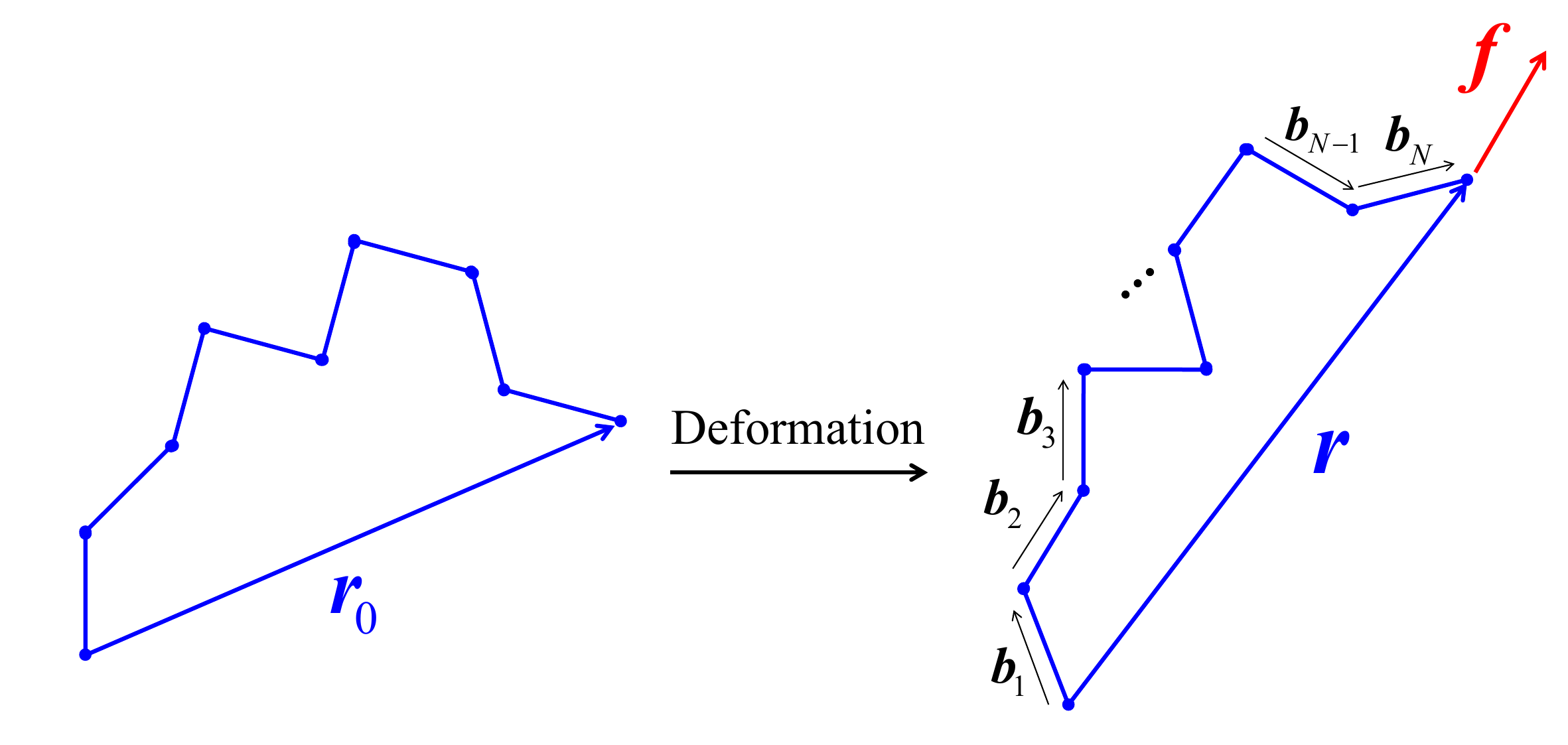}
	\end{minipage}		
}
	\subfigure[] 
{
	\begin{minipage}{7.5cm}
	\centering       
	\includegraphics[width=75.00mm,height=35.00mm]{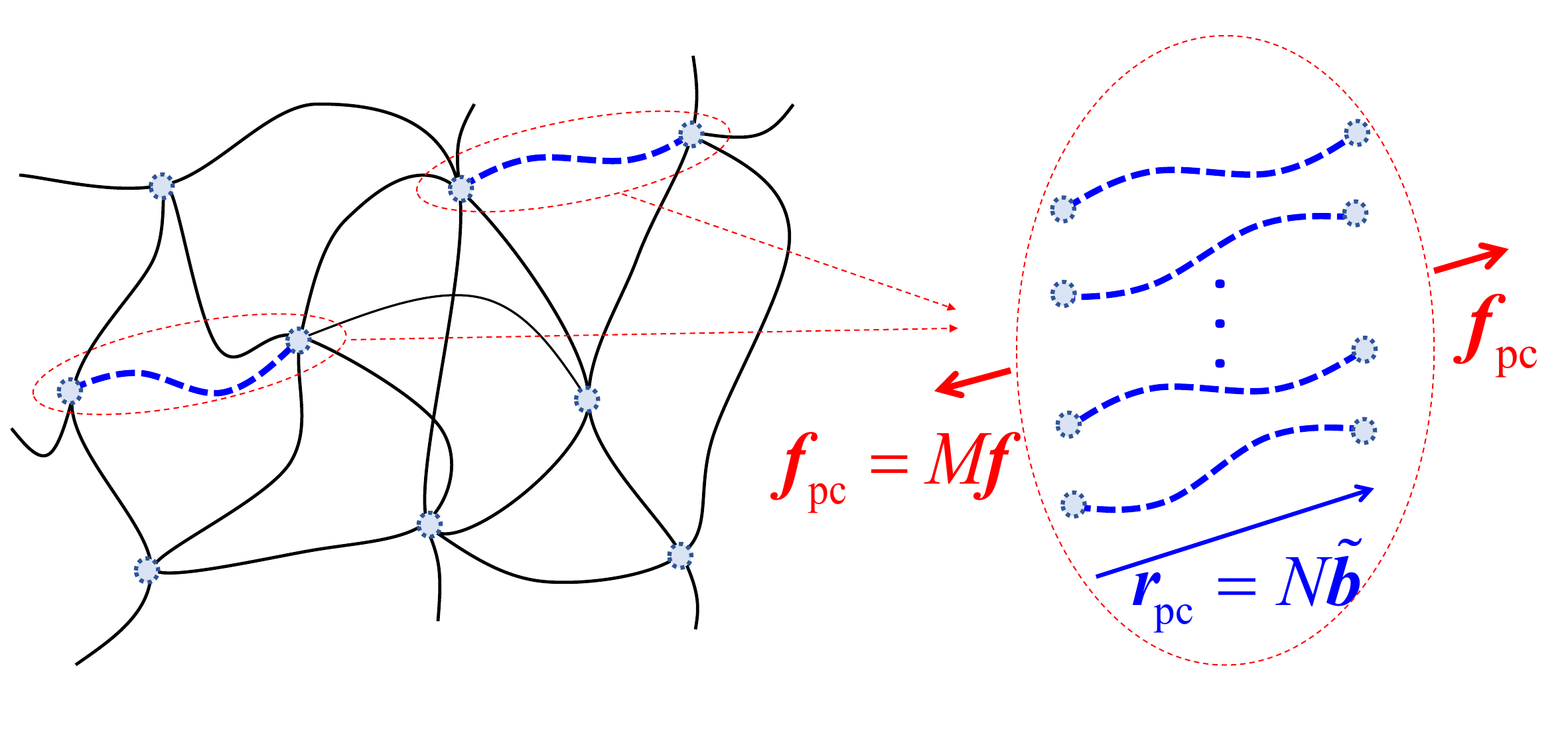}
\end{minipage}		
}
\caption{(a) A sketch for the freely jointed chain subject to external force; (b) a sketch for the network model:  the average number of chains with end-to-end vector $\vect r_\text{pc}=\vect r$ (dashed lines) is $M$ and the total force is $\vect f_\text{pc}=M\vect f$.} 
	\label{fig:chainmodel} 
\end{figure}

Next, we consider a chain subjected to an external force $\vect f$, as illustrated in Fig. \ref{fig:chainmodel}a. The chain has a rotational potential
\begin{equation}
H_r\left(\{\vect b_i\}\right)=-\vect f\cdot\vect r=-\sum_{i=1}^{N}\vect f\cdot\vect b_i=\sum_{i=1}^{N}H_b(\vect b_i),
\end{equation}
where $H_b(\vect b_i)= -\vect f\cdot\vect b_i$ is the potential energy of the $i$-th segment. 

\textcolor{black}{Due to thermal motion, the possible configuration $\vect b$ for each segment $\vect b_i$ can be treated as ergodic on a sphere surface with the constraint $||\vect b||=b$.} The probability of a segment in the configuration of $\vect b$ is given as 
\begin{equation}
	\label{segment probability}
	p(\vect b) =\dfrac{1}{Z}e^{\vect f\cdot\vect b/k_BT},
\end{equation}
with $k_B$ and $T$ being the Boltzmann's constant and absolute temperature, respectively. The above function is an intergration  Here, $Z$ is the partition function that can be calculated as  
\begin{equation}
	\label{segment partition}
Z=\int_{|\vect b|=b} e^{\vect f\cdot\vect b/k_BT}\text d\vect b=4\pi b^2\frac{\sinh\left(\xi\right)}{\xi},
\end{equation}
Here $\xi$ is defined by $\xi=fb/k_BT$ with \textcolor{black}{$f=||\vect f||$}.

The Gibbs energy of a segment is given by
\begin{equation}
	\label{gibbs free energy}
	G(f)=-k_BT\ln Z=k_BT\ln \frac{\xi}{\sinh\xi},
\end{equation}
where the constant term is neglected. 

The average of $\vect b$, denoted as  $\tilde{\vect b}=\left<\vect b\right>$, is expressed as
\begin{equation}
	\label{distance force}
\tilde{\vect b}= -\frac{\partial G}{\partial \vect f}= \frac{1}{Z} \int_{|\vect b|=b} \vect be^{\vect f\cdot\vect b/k_BT}\text d\vect b=b\frac{\vect f}{f}\mathcal L\left(\xi\right),
\end{equation}
where $\mathcal L(x)=\coth x-1/x$ is the Langevin function.  The inverse form of the above equation is written as
\begin{equation}
	\label{force distance}
	\vect f=\dfrac{k_BT}{b}\frac{\tilde{\vect b}}{\tilde{ b}} \mathcal L^{-1}\left(\frac{\tilde{b}}{b}\right)
\end{equation}
with $\tilde{b}=\vert\tilde{\vect b}\vert$. 

Using Legendre transformation, we obtain the Helmholtz energy of the segment:
\begin{equation}
	\label{helmholtz free energy}
	F(\tilde b)=G+\frac{\partial G}{\partial \vect f}\cdot \vect f=k_BT\left[\frac{\tilde b\beta}{b}+\ln\frac{\beta}{\sinh\beta}\right],
\end{equation}
with $\beta=\mathcal L^{-1}\left(\tilde{b}/b\right)$. 

In the above discussion, the chain is analyzed through an isotensional ensemble of segments by fixing the external force. Equivalent formulations can be obtained in the thermodynamic limit by considering the chain as fixed in its end-to-end position $\vect r$, or equivalently, fixed $\tilde{\vect b}$ \cite{winkler2010equivalence,weiner2012statistical,manca2012elasticity,buche2020statistical}. In this case, the chain is treated as an isometric ensemble of segments. For both ensembles, a chain comprising $N$ segments is characterized with the energies $G_\text{chain}=NG$ and $F_\text{chain}=NF$. The force and displacement are specified as $f_\text{chain}=f=f_\text{segment}$ and $r_\text{chain}=r=N\tilde{ b}$ due to the series connection of segments. This fact indicates that $f$ and $\tilde b$ serve as the intensive and extensive quantities, respectively. It is worth noting that $G$, $F$, and $\tilde{b}$ defined above represent time-averaged quantities for a single segment over all its possible configurations. However, in a system with a large number of segment replicas, these quantities can be regarded as macroscopic thermodynamic properties per segment \cite{huang2008statistical}. This statistical principle is admitted throughout the work without further specifications.
 
We now consider a cross-linked polymer network where the ends of each chain are attached to cross-linking points, as shown in Fig. \ref{fig:chainmodel} (b). In a small volume element $V$, let $M$ denotes the number of chains with the same end-to-end vector $\vect r$. Since it is highly likely that these $M$ chains are attached to different linkers, they can be modeled as a system of parallel chains with the end-to-end vector $\vect r$, where the total force contributed by the chains is additive. 

In a parallel chain system, the chain length is $r_\text{pc}=r=N \cdot \tilde b$ and the total conservative force is $f_\text{pc}=M\cdot f$. Moreover, the Helmholtz energy is given by $F_\text{pc}=M \cdot F_\text{chain}=MN\cdot F(\tilde b)$. However, a system of $M$ parallel chains cannot be described in the same manner as a single chain. Specifically, neither the $f$ nor the $\tilde b$ can be regarded as the intensive quantity that is independent of the segment number. Accordingly, none of them can be treated as the extensive quantity that is proportional to the number of segments. The difference between a parallel-chain system and a single chain lies in the arrangement of segments: the former involves a hybrid pattern with both series and parallel connections, while the latter only consists of series connections.  

This difficulty can be resolved by using another thermodynamic conjugate pair. Noting the relation of $f_\text{pc}r_\text{pc}=MN \cdot f\tilde b$, it appears that the system of $M$ parallel chains can be described in a similar fashion as a single chain if we treat $f\tilde b$ as a new generalized force. Defining the relative stretch as $\lambda=r/r_0=\sqrt N\tilde b/b$, the conjugate variable to $f\tilde b$ turns out to be $\ln\lambda$, using the relation that $\text dF=f\text d\tilde b=f\tilde b\text d(\ln\lambda)$. Then the generalized force $f_\text{pc}r_\text{pc} $ and free energy $F_\text{pc}$ for the parallel chains are unified as $MN\cdot f\tilde b$ and $MN\cdot  F$, respectively. The logarithmic stretch for the parallel-chains and the segments are both $\ln\lambda$. Hence, the new pair of variables $(f\tilde b,\ln\lambda)$ is employed as extensive and intensive quantities. 

The Helmholtz energy of a segment in the parallel-chain system, described by $ \ln\lambda$, is obtained from Eq. \eqref{helmholtz free energy} by replacing $\tilde b/b$ with $\lambda/\sqrt N$, as $b$ and $N$ are both material constants. The potential energy of a segment in the state of $f\tilde b$ is determined by the Legendre transformation to $F(\ln\lambda)$:
\begin{equation}
	\label{augmented legendre}
	\tilde G=F-f\tilde b\ln\lambda=F-\dfrac{\partial F}{\partial \ln\lambda}\ln\lambda.
\end{equation}
The function $\tilde G$ serves as a new Hamiltonian of segments subjected to the generalized force $f\tilde b$ for either a chain or a parallel-chain system. It is a minimum of the function $\ln\lambda $ with respect to $f\tilde b$.

Let $n$ denote the number of chains in the volume $V$, making the number of parallel-chain systems $n/M$. Under deformation, both $n$ and $V$ remain conserved if the region is assumed to be volume-preserving \cite{treloar1975physics,xing2007thermal}. The polymer network in $V$ can then be modeled as a system with $nN$ segments. These segments are independent but allow weak interactions to achieve equilibrium. Since the only distinguishable quantity of a segment is its spatial direction $\vect u\in \mathbb{S}^2$, the energy state of a segment is identified by the Hamiltonian $\tilde G(\ln\lambda)=\tilde G(\vect u)$. This implies the relative stretch also possesses a dependence on the spatial direction, namely $ \lambda=\lambda(\vect u)$.  Then the probability of a segment occupying a state with potential $\tilde G$, corresponding to the direction $\vect u$, is given by
\begin{equation}
	\label{partition}
	P(\vect u)=\frac{1}{\tilde Z}e^{-\tilde G/k_BT}, \quad \tilde Z=\int_{|\vect u|=1} e^{-\tilde G/k_BT}\text d\vect u,
\end{equation}
where $\tilde Z$ is the partition function. \textcolor{black}{It should be noted that $\vect u$ is actually an average of the segment direction in an equilibrium chain. Hence, it represents the chain's orientation.}

The augmented free energy per unit volume corresponding to $\tilde G$ is 
\begin{equation}
	\label{augmented free energy}
	\mathcal G=-\rho N\cdot k_BT\ln \tilde Z=\rho N\cdot\left<\tilde G\right>-\rho N\cdot T\tilde S,
\end{equation}
where $\rho=n/V$ is the number of chains per unit volume. The entropy can be further calculated as $ \tilde S=-k_B\left<\ln P\right>$. The $\mathcal G$ turns out to be the Gibbs energy density of the network, as will be shown latter.

The continuum elastic energy density of the network is expressed as the average Helmholtz energy density:
\begin{equation}
	\label{elastic free energy}
W=\rho N\cdot\left<F\right>=\rho N\int_{|\vect u|=1} P\cdot F(\ln \lambda)\text d\vect u.
\end{equation}

The elastic energy density is also identified as $W=W(\vect F)$ with $\vect F$ being the deformation gradient defined by $\vect F={\partial \vect x}/{\partial\vect X}$. Here, the deformation gradient maps line element $\text d \mathbf X$ in the reference configuration to $\text d \mathbf x$ in the deformed configuration. We then infer that the relative stretch should be determined through $\lambda = \lambda(\vect u, \vect F)$. 
Thus, the free energy $W=W(\ln\lambda)$ in Eq. \eqref{elastic free energy} turns out to be a functional of the function $\lambda=\lambda(\vect u, \vect F)$. 

Based on the virial theorem \cite{tsai1979virial,subramaniyan2008continuum,admal2010unified,bobrov2010virial}, the Cauchy stress tensor $\vect\sigma$ at thermodynamic equilibrium is expressed as
\begin{equation}
	\label{cauchy stress}
	\vect\sigma=\rho \left<\vect f_\text{pc}\otimes \vect r_\text{pc}\right>-\rho Nm\left<\vect v\otimes\vect v\right>=\rho N \left<f\tilde b\vect u\otimes\vect u\right>-p\vect I,
\end{equation}
where the first term represents the elastic potential between cross-linking points, and the second term accounts for the translation kinetic energy of the mass center of the segments. Here, $m$ and $\vect v$ denote the segmental mass and velocity, respectively. Since the translation velocity is isotropic, the kinetic part eventually reduces to an isotropic pressure, denoted as $p\vect I$.

As emphasized in \citep{treloar1975physics,xing2007thermal,xing2011biaxial}, the cross-linking junctions undergo thermal (micro-Brownian) fluctuations coupled with the motion of the attached chain segments. This leads to statistical fluctuations of relative positions between junctions, namely the chain configuration. Since the chain stretch is directly related to the orientation, such statistical fluctuations is then described by the orientation probability function, i.e. the Eq. (\ref{partition}) in the current theory. The key novelty of our approach lies in the consideration of the average behavior of the chain force-configuration dyad $ \mathbf{f} \otimes \mathbf{r} $, whose ensemble average directly enters the continuum Cauchy stress, as expressed in Eq. (\ref{cauchy stress}). Such consideration arises form our attempt to eliminate the thermodynamic inconsistency of the conjugate pairs in parallel chain systems. However, the arguments of James and Guth \cite{james1943theory} regarding this issue, as discussed by Treloar \cite{treloar1975physics}, mostly focus on the average chain force, which may not be easily linked to a macroscopic observable.

Equation \eqref{cauchy stress} shows that the Cauchy stress is measured as the average of fluctuating $f\tilde b\vect u\otimes \vect u$. According to continuum thermodynamics, the Cauchy stress is also given by $\vect\sigma=\partial W/\partial\vect h-p^*\vect I$ with $\vect h=\ln\left(\vect F\cdot\vect F^T\right)/2$ being the Eulerian logarithmic strain tensor. This is equivalent with the static equilibrium condition that the Legendre transformation $W-\vect\sigma:\vect h$ attains a minimum with respect to the function $\vect h=\vect h(\vect\sigma)$. The incompressible condition is expressed as $\vect h:\vect I=\text {tr}\vect h=0$, which eliminates the isotropic pressure $p^*$ and the Lagrangian multiplier $p$ in the term $\vect\sigma:\vect h$. Using Eqs. \eqref{helmholtz free energy}-\eqref{partition} and \eqref{cauchy stress}, the minimum condition in the isothermal case becomes
\begin{equation}
	\label{minimum condition}
\delta\left( W-\vect\sigma:\vect h\right)=\rho N\int Pf\tilde b\left[\frac{\partial\ln\lambda}{\partial\vect h}-\vect u\otimes\vect u\right]\delta\vect h\text d\vect u=0,
\end{equation}
which requires 
\begin{equation}
	\label{lambda u}	
	\ln\lambda=\vect h:\vect u\otimes\vect u,
\end{equation} 
as a sufficient condition.

The above relation establishes a direct connection between chain stretch and macroscopic deformation measures, namely, the micro-macro transition. It is noted that previous works \citep{amores2021network, zhan2023new} have also developed a similar non-affine mapping
$\lambda_n = \mathbf{n} \cdot \mathbf{U} \cdot \mathbf{n}$,
where $ \mathbf{U} $ is the right stretch tensor and $ \mathbf{n} $ denotes the orientation of a specific chain in the undeformed state. This Lagrangian mapping assumes that the stretch of a given chain is deterministically governed by its initial orientation and the applied macroscopic deformation. In the current theory, however, a chain may orient to any direction due to thermal fluctuations. Hence, the kinematic behavior of a specified chain is inherently non-deterministic since the orientating direction $\mathbf u$ is not deterministic. As such, our current formulation describes the chain kinematics and orientation probability as spatial direction-field quantities, thereby establishing a Eulerian framework. The two mappings are also mathematically different. Even if we assume a deterministic rotation $\mathbf{u} = \mathbf{R} \mathbf{n}$ for the chain orientation, one cannot derive
$\ln \lambda = \mathbf{R} \mathbf{n} \cdot \ln(\mathbf{R} \mathbf{U} \mathbf{R}^T) \cdot \mathbf{R} \mathbf{n}$
directly from
$\lambda = \mathbf{n} \mathbf{U} \mathbf{n}$, and vice versa. As will be demonstrated in Appendix A, the current model also offers enhanced predictive performance than the previous work. Nevertheless, the earlier model may still be advantageous in certain engineering contexts due to its simplicity: it does not require explicit treatment of orientation anisotropy and can yield closed-form expressions in the small to moderate strain regime, making it attractive for applications where computational efficiency is prioritized.

Classical microstructurally motivated hyperelastic models, such as those cited in \cite{miehe2004micro}-\cite{zhan2023new}, are primarily developed within a Lagrangian framework, where the chain stretch/force is treated as a deterministic function of its initial orientation and the macroscopic deformation/stress. These models often assume fixed or isotropic orientation distributions and neglect the role of thermal fluctuations in reorienting chains. While in our theory, the behavior of polymer chains is described jointly by two aspects: the stretch of a chain aligned with a spatial direction $ \mathbf{u} $ is determined by the kinematic relation $ \ln \lambda_u = \mathbf{u} \cdot \mathbf{h} \cdot \mathbf{u} $, and the probability of observing a chain in that direction is given by a probability density function $P(\mathbf u)$ derived from statistical mechanics, which is conceptually distinct from the Lagrangian models. This approach provides a new and physically grounded perspective on the microstructural modeling of polymer networks.

By taking statistical average of Eq. \eqref{augmented legendre}, we obtain $\rho N\left<\tilde G\right>=W-\vect\sigma:\vect h$, indicating that the statistical average of the new defined chain potential $\tilde G$ corresponds to the thermodynamic enthalpy of the continuum polymer. This enthalpy is also referred to as the complementary elastic energy in continuum mechanics.  Consequently, the augmented free energy in Eq. \eqref{augmented free energy} can be expressed as $\mathcal G(\vect\sigma, T)=W-\vect\sigma:\vect h-T\tilde S$, representing the thermodynamic Gibbs energy of the polymer network.  The physical picture will be more clear if we treat the elastic energy  $W$ as an internal-energy-like quantity. Given stress and temperature, the thermodynamic equilibrium requires $\delta\mathcal G=0$, which has already been satisfied by the relation $\partial\mathcal G/{\partial P}=0$ through Eq. \eqref{partition}.

\textcolor{black}{For better clarity, the theoretical framework is summarized in Table 1.} As shown, the probability of a segment orientating to direction $\vect u$, eventually expressed as a function of continuum deformation,  is independent of the temperature. However, one might expect thermal motion to favor isotropic configurations at elevated temperatures, seemingly contradicting the theory. This is not the case. At elevated temperatures, larger stress is required to achieve the same strain, implying that more energy is needed to sustain a specific orientation under a given deformation.

\begin{tcolorbox}[title= Table 1. A flow chart for the hyperelastic theory, colframe=black, coltitle=black,colback=white, colbacktitle=white, fonttitle=\bfseries,boxrule=0.5pt ]
	\setstretch{1.0}{
1. Kinematics:
		\[ \vect F=\frac{\partial \vect x}{\partial\vect X},\quad \vect h=\frac{\ln\left(\vect F\cdot\vect F^T\right)}{2}.\]
2. Micro-macro mapping:
		\[\lambda(\vect u)=e^{\vect h:\vect u\otimes\vect u}.\]
3. Helmholtz energy of a segment:
	\[	F(\lambda)=k_BT\left[\frac{\lambda\beta}{\sqrt N}+\ln\frac{\beta}{\sinh\beta}\right], \beta=\mathcal L^{-1}\left(\frac{\lambda}{\sqrt N}\right).\]
4. Probability of a segment/chain in the direction $\vect u$:
	\[P(\vect u)=\frac{e^{-\tilde G/k_BT}}{\int_{|\vect u|=1} e^{-\tilde G/k_BT}\text d\vect u},\]
	\[\tilde G=F-\dfrac{\partial F}{\partial \lambda}\lambda\ln\lambda.\]
5. Continuum stress-strain relation:
	\[W=\rho N\int_{|\vect u|=1} P(\vect u)\cdot F(\lambda)\text d\vect u,\]
	\[\vect\sigma=\dfrac{\partial W}{\partial\vect h}+p\vect I.\]
	}
\end{tcolorbox}

To quantify chain anisotropy, we can define an order parameter tensor \cite{de1971short} as $\vect A=\left<\vect u\otimes\vect u\right>-\frac{1}{3}\vect I$. The alignment to a specified direction $\vect n$ can be characterized by a scalar order parameter as
\begin{equation}
	\label{lambda u}	
 A_n=\frac{3}{2}\vect n\cdot\vect A\cdot\vect n=\frac{3}{2}\int_{|\vect u|=1} P(\vect u)\cdot (\vect u\cdot\vect n)^2\text d\vect u-\frac{1}{2}. 
\end{equation} 
The bounds of $A_n$ are $-1/2\leq A_n\leq1$, where $A_n=1$ indicates perfect alignment along $\vect n$, $\vect A=\vect 0$ corresponds to isotropic states, and $A_n=-1/2$ implies all segments are perpendicular to $\vect n$. 

\begin{figure}[]
		\begin{minipage}{8cm}
			\centering       
			\includegraphics[width=80.00mm,height=40.00mm]{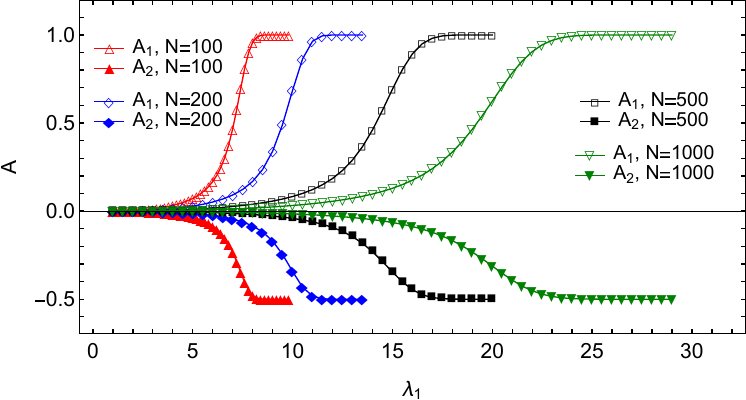}
		\end{minipage}
	\caption{Chain anisotropy in  uniaxial tension with order parameter $A_1$ in the stretched direction and $A_2$ in lateral direction for different values of chain segment $N$.} 
	\label{fig:order} 
\end{figure}

Figure \ref{fig:order} shows the order parameter for uniaxial tension with different chain lengths. Here, $\lambda_1$ is the continuum principal stretch in the stretched direction, $A_1$ and $A_2$ are the order parameter in the stretched and lateral directions, respectively. The results show that longer chains induce less anisotropy, as orientation behavior is governed by segmental energy that is inversely proportional to chain length. When the condition $A_1=1$ is achieved, all segments align to the stretched direction. 

It should be notice that in the large strain region, chains in the primary stretched direction can be fully extended, corresponding to the limiting case of rigid segment model. Under such conditions, a modified chain model considering the bond extensibility may be needed \cite{netz2001strongly,manca2012elasticity,mao2017rupture,buche2020statistical,buche2022freely,mulderrig2023statistical}. Such mechanism is of crucial importance to account for the damage behavior\cite{wang2019quantitative} and can be incorperated in the micro-sphere framework using affine and non-affine approaches\citep{buche2021chain,mulderrig2021affine}. Based thereon, the current theory with microsphere structure and non-uniform chain density may be extended to include bond extensibility by modifying the orientation Hamiltonian to account for internal energy contributions. Moreover, recent studies on polydisperse networks\citep{xiao2021modeling,araujo2024micromechanical}, highlight that chain stretch and rupture behavior are significantly influenced by the chain length distribution. Shorter chains tend to reach their extensibility limits earlier, leading to localized damage or scission.  While the current theory assumes a uniform chain length for simplicity, it may be extended to accommodate polydispersity by reformulating the statistical mechanics for chains of specified length and integrating their contributions across the full length distribution. These prior works offer a promising path for our future development of enabling a more accurate description of network fracture and damage evolution in realistic materials.  As is demonstrated in our previous studies \citep{zhan2023general,zhan2024review}, the predictive capability of damage models under complex loading paths depends on both the chain-level damage mechanism and the micro-macro mapping. Given that the current model demonstrates superior predictive performance compared to traditional mapping approaches, we infer that a damage model incorporating both chain extensibility and the present theory may further enhance accuracy under multiaxial loading conditions.

\begin{figure}[]
	\subfigure[] 
	{
		\begin{minipage}{4cm}
			\centering       
			\includegraphics[width=40.00mm,height=38.00mm]{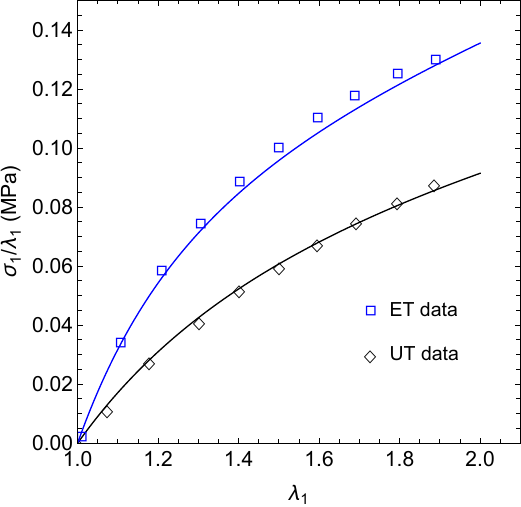}
		\end{minipage}
	}
	\subfigure[] 
	{
		\begin{minipage}{4cm}
			\centering       
			\includegraphics[width=40.00mm,height=38.00mm]{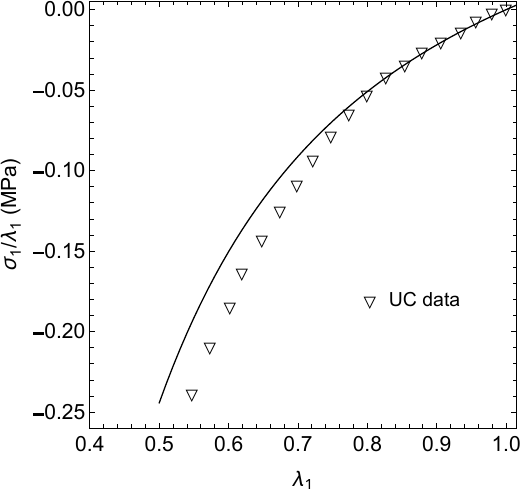}
		\end{minipage}
	}
	\subfigure[] 
	{
		\begin{minipage}{4cm}
			\centering       
			\includegraphics[width=40.00mm,height=38.00mm]{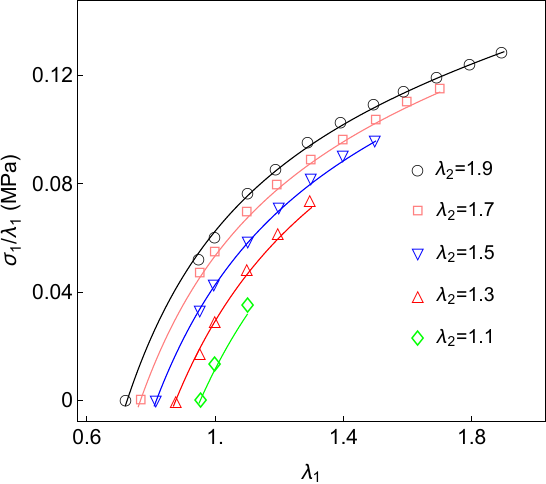}
		\end{minipage}
	}
	\subfigure[] 
	{
		\begin{minipage}{4cm}
			\centering       
			\includegraphics[width=40.00mm,height=38.00mm]{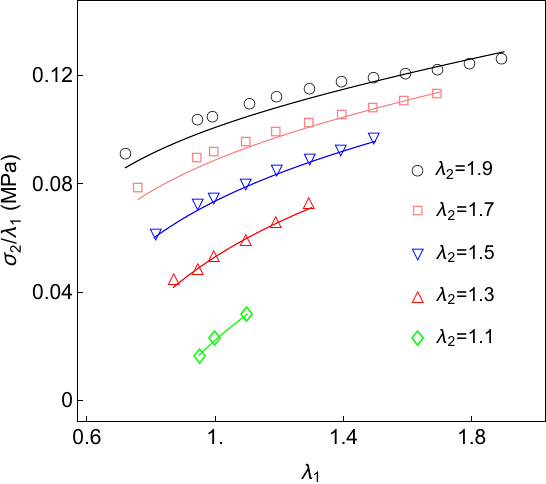}
		\end{minipage}
	}
	\caption{Comparing the  experimental data \cite{kawamura2001multiaxial} and predicted stress-stretch responses in (a) uniaxial tension (UT, tension in direction $\vect{e}_1$ with the principal stretch $\lambda_1$ in the loading direction) and equibiaxial tension (ET, tension in two perpendicular directions $\vect e_1$ and $\vect e_2$ with equal stretch $\lambda_2=\lambda_1$), and (b) uniaxial compression (UC, compression in direction $\vect e_1$), and  biaxial responses for stress in (c) the stretched direction $\vect e_1$ and (d) the fixed direction $\vect e_2$, with $\rho k_BT=0.160$ MPa. In the biaxial tests, a set of specimens were first equibiaxially loaded to different maximum stretches in directions $\vect e_1$ and $\vect e_2$. Then the stretch $\lambda_1$ was gradually reduced until the stress $\sigma_1$ vanished, while keeping the stretch ratio $\lambda_2$ unchanged. } 
	\label{fig:pdms} 
\end{figure}

\begin{figure}[]
	\begin{minipage}{7.0cm}
		\centering       
		\includegraphics[width=60.00mm,height=62.00mm]{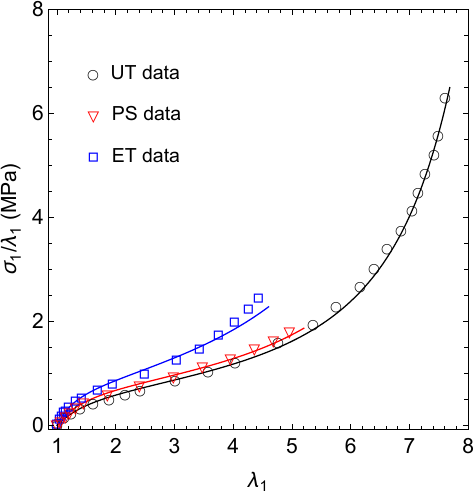}
	\end{minipage}		

	\caption{Comparing the predicted stress-stretch responses in UT, ET and PS (pure shear, tension in direction $\vect e_1$ while fixing the stretch ratio  $\lambda_2=1$ in the perpendicular direction) of Treloar's data \cite{treloar1944stress} using $\rho k_BT=0.99$ MPa and $N=146$.  } 
	\label{fig:treloar} 
\end{figure}

For small to moderate deformations or sufficiently large $N$, the condition $\lambda/\sqrt N\ll1$ holds. In this case the elastic energy of chains is reduced to the Gaussian form and the segmental energy is given by $\tilde G=F-\brac{{\partial F}/{\partial\ln\lambda}}\ln\lambda=3k_BT\left[\lambda/\sqrt N\right]^2\left[1/2+\ln\lambda\right]\approx 0$. The segments then exhibit nearly isotropic orientation. \textcolor{black}{Following the flow chart in Table 1}, the predicted stress-deformation responses are compared with experimental data \cite{kawamura2001multiaxial}, including uniaxial tension, compression and biaxial tension, as presented in Fig. \ref{fig:pdms}. For large deformations, chain segments follow an anisotropic orientation and the theoretically predicted stress responses are compared with the classical Treloar's data \cite{treloar1944stress}, as shown in Fig. \ref{fig:treloar}. In both cases, the proposed theory demonstrates significantly improved performance, relying on minimal and well-defined physical parameters, which surpasses the prediction capability of classic hyperelastic models reported in the literature \cite{dal2021performance,zhan2023new}. It is noted that several successful phenomenonlogical models also involve with the logarithmic forms\cite{gent1996new,kulcu2020hyperelastic,anssari2022assessment}. This interesting issues will further investigate in future studies.  A further quantative analysis of the model's predictive is provided in the Appendix A.

It is important to note that the orientation behavior of a chain is governed by the segment level energy $\tilde G$, rather than the chain-level energy $N\tilde G$, as shown in Eq. (\ref{partition}). As the $\tilde G$ decreases with increasing chain length $N$, the chain anisotropy then decreases. In the limiting case, a chain with an infinite number of segments would remain isotropic, as it cannot sustain a shear force under finite deformation, similar to a polymer solution.

Bridging molecular models with continuum-scale behavior has long been a central challenge in multiscale modeling of materials. In crystalline materials, molecular kinematics are typically constrained to small oscillations around equilibrium positions since the thermal agitation is much weaker than the constraints imposed by crystal lattice. In such cases, molecular can be regarded as fixed in the geometrical background and continuum kinematics, such as the affine mapping, can reasonably approximate molecular motion. However, for amorphous soft materials without strong crystal lattice constraints, molecular configurations are largely governed by thermal fluctuations, and their positions cannot be described deterministically. In these systems, the molecular response must be modeled using statistical theories that account for the probability of a molecule occupying a given position/configuration. In fact, when thermal effects are sufficiently strong, the resulting continuum fields in both soft or hard materials should be determined through statistical average with non-uniform probability distributions, rather than traditional homogenization procedure with uniform probability. This perspective has also been emphasized in the field of molecular dynamics simulation \citep{tadmor2011modeling}. The present theory is built upon this statistical foundation. The proposed micro-macro mapping and probability density function may be extended to a wider class of soft materials in which the interactions among chains or segments are relatively weak, since the theoretical framework is derived from the standard weak-interaction statistical ensemble theory. For more complex materials, such as highly entangled polymers, more advanced statistical approaches involving with strong interactions would likely be necessary.

In conclusion, we have developed a statistical theory that relates chain kinematics, including chain stretch and orientation, to continuum deformation. A new Hamiltonian for the chain segment enables the polymer network to be described as an ensemble of segments in a unified form. By using the equivalence between the Cauchy stress and the virial stress, a novel relation is established that connects the chain stretch with spatial direction and the continuum Eulerian logarithmic strain. The orientation distribution is then directly obtained through the statistical probability function. The predicted mechanical responses outperform those using the classic hyperelastic models, relying on only one or two physically grounded parameters. This theory can deep our understanding on the microscopic kinematics of polymer chains in the elastic condition and can also be extended to many other cases, such as damage effects and multi-field coupling responses. 

\hfill
\begin{acknowledgments}
This work is supported by the National Natural Science Foundation of China (Grant Nos. 12202378, 12321002), the Postdoctoral Fellowship Program of CPSF (Grant No. GZC20230963), and the 111 Project (Grant No. B21034).
\end{acknowledgments}

\section*{Appendix A}
This appendix provides quantitative comparisons between the current model and other classical hyperelastic models to demonstrate the significant improvement of the predictive ability.

We first calibrate the model parameters strictly using the metric of average relative error (ARE), defined as \citep{zhan2023new}
	\begin{equation}
		\label{eq:are}
		\text{ARE} = \dfrac{1}{n} \sum_{i=1}^{n} \left| \dfrac{S_{\text{simu}}(\lambda_i) - S_{\text{exp}}(\lambda_i)}{S_{\text{exp}}(\lambda_i)} \right|,
	\end{equation}
where $ n $ is the number of data points, and $ S_{\text{simu}}(\lambda_i) $ and $ S_{\text{exp}}(\lambda_i) $ are the simulated and experimental stresses at a given stretch $ \lambda_i $, respectively. The results are presented in Tab. \ref{tab:are}. All model parameters are calibrated by minimizing the ARE with respect to uniaxial tension (UT) data, and the predictions are then made for biaxial deformation modes.

For Treloar's data, we compare the performance of our model with the Biot-chain model, the affine model, the eight-chain model, and the equilibrated chain model. The stress–strain curves for the four competing models can be found in \cite{zhan2023new} and are not repeated here to avoid redundancy. To the best of our knowledge, these models represent the class of physically based two-parameter models that are capable of capturing large deformation features. In Tab.\ref{tab:are}, the parameters of the four competing models are reproduced from \cite{zhan2023new} and converted to be expressed by $ \rho k_BT $. The results show that our current model achieves the lowest global ARE for Treloar’s data, indicating the highest predictive ability.

For the PDMS data within moderate stain range, we compare the current model(isotropic Gaussian form) with the one-parameter Biot-chain model and the neo-Hookean model. In fact, the neo-Hookean model is a limiting case of the affine, eight-chain, and equilibrated chain models in the small-to-moderate deformation regime \citep{zhan2023new}. Again, the results indicate that our model provides more accurate predictions for biaxial tests.

\begin{table*}[]
	\caption{The average relative error (ARE) of models predicting Treloar's data and PDMS data.}
	\centering
	\begin{tabular}{cccc}
		\hline
		\\
		\qquad\textbf{Materials}\qquad &\qquad\textbf{Models}\qquad\qquad & \qquad\makecell{\textbf{Parameters;}\\ $\rho k_BT$/MPa} \qquad\qquad & \textbf{ARE}(\%)  \\
		\\
		\hline
		\\
		\multirow{12}{*}{\makecell{Vulcanized \\ rubber \citep{treloar1944stress}}}&\textbf{Current model} & \makecell{$\rho k_BT=0.99$;\\$N=146$}&  \makecell{UT: 6.27;\quad PS: 6.31;\\ \quad ET: 5.56; \quad Total: \textbf{6.05}} \\
		\\
		&Biot-chain model \cite{zhan2023new}& \makecell{$\rho k_BT=0.50$;\\$N=61.92$}& \makecell{UT: 8.49;\quad PS: 5.24; \\ \quad ET: 8.58; \quad Total: 7.44} \\
		\\
		&Affine model & \makecell{$\rho k_BT=0.31$;\\$N=62.3$}&   \makecell{UT: 7.89;\quad PS: 7.08;\\ \quad ET: 23.21; \quad Total: 12.73 } \\
		\\
		&Eight-chain model & \makecell{$\rho k_BT=0.28$;\\$N=26.15$}& \makecell{UT: 6.38;\quad PS: 9.26; \\ \quad ET: 23.80; \quad Total: 13.15 } \\
		\\
		&Equilibrated model & \makecell{$\rho k_BT=0.27$;\\$N=31.13$}& \makecell{ UT: 8.10;\quad PS: 11.65;\\ \quad ET: 29.94; \quad Total: 16.56}\\
		\\
		\hline			
		\\
		\multirow{5}{*}{\makecell{PDMS \citep{kawamura2001multiaxial}}} & \textbf{Current model} & \makecell{$\rho k_BT=0.160$} &  UT: 3.54,\quad $\sigma_1$: \textbf{2.42},\quad $\sigma_2$: \textbf{2.23} \\
		\\
		& Biot-chain model \cite{zhan2023new}& \makecell{$\rho k_BT=0.091$}& UT: 2.12,\quad $\sigma_1$: 6.69,\quad $\sigma_2$: 2.34 \\
		\\
		&neo-Hookean model& \makecell{$\rho k_BT=0.061$} &   UT: 2.51,\quad $\sigma_1$: 20.32,\quad $\sigma_2$: 6.59 \\
		\hline
	\end{tabular}
	\label{tab:are} 
\end{table*}

In it noted that the Biot-chain model \citep{zhan2023new} using the non-affine mapping $\lambda = \mathbf{n} \mathbf{U} \mathbf{n}$ performs better than the affine models. This can be explained by the randomness of the chain orientation behavior in the current theory. Consider a group of chains initially aligned along a specific direction $ \mathbf{n} $; under deformation, these chains can diffuse into all spatial directions with non-uniform probability due to thermal fluctuations. Therefore, a proper description of the chain behavior should be constructed within the Eulerian frame work, which tracks the kinematics of a chain in the specified spatial direction and do not intend for telling which chain is in this direction. If an effective stretch associated with the initial direction $ \mathbf{n} $ need be defined, it should account for contributions from chains oriented in all directions in the current configuration. In affine models, chains initially aligned along $ \text d\mathbf{X}$ direction are assumed to undergo identical stretch and rotation to an identical configuration $\mathbf F\text d\mathbf{X}$, without considering the redistribution of orientation caused by thermal motion. The non-affine mapping $\lambda = \mathbf{n} \cdot \mathbf{U} \cdot \mathbf{n}$ is derived as the average value of the affine deformation in all directions projecting into the direction $\bm n$, and thus represents an equivalent stretch\citep{zhan2023new}. This might be the reason why it has satisfactory performance. However, it is derived through an equal-probability averaging procedure and does not capture the statistical averaging over reoriented configurations that naturally occurs in real networks. As a result, it cannot fully reproduce the predictive performance of the present model. We infer that other non-affine models may also benefit from similar considerations, either explicitly or implicitly, through their specific assumptions and formulations.

To further support our claim regarding predictive superiority, we additionally refer to the comprehensive study of \cite{dal2021performance}, in which the performance of 44 existing hyperelastic models in fitting Treloar's experimental data was systematically evaluated.

In alignment with the methodology of \cite{dal2021performance}, we adopt the same parameter calibration procedure. Specifically, model parameters are calibrated by minimizing the total squared error with respect to uniaxial tension (UT) data:
\begin{equation}
	\label{eq:se}
	\text{Squared Error} = \sum_{i=1}^{n} \left[ S_{\text{simu}}(\lambda_i) - S_{\text{exp}}(\lambda_i) \right]^2.
\end{equation}
And comparison is then performed using the relative squared error (RSE), defined as
\begin{equation}
	\label{eq:rse}
	\text{RSE} = \sum_{i=1}^{n} \dfrac{ \left[ S_{\text{simu}}(\lambda_i) - S_{\text{exp}}(\lambda_i) \right]^2 }{ S_{\text{exp}}(\lambda_i) }.
\end{equation}

The corresponding results are shown in Tab. \ref{tab:rse}. According to the results of ``UT only fit (Treloar)" in Tab. 5-26 of \cite{dal2021performance}, our model achieves the 5th-best predictive performance among all evaluated 44 models. Notably, the top four models that outperform ours all involve three parameters, relying on additional physical or phenomenological assumptions. In contrast, our model attains comparable accuracy using only two parameters, and is derived entirely from a statistically grounded framework. 

These results highlight the practical efficacy and theoretical robustness of our theory. However, due to the incorporation of microsphere integration to account for chain orientation anisotropy, the computational cost is generally higher than that of purely analytical models in the literature.

\begin{table*}[htbp]
\caption{The relative squared error (RSE) of models predicting Trealoar's data.}
\centering
\begin{tabular}{ccccc}
	\hline
	\\
	\makecell{Rank of performance\\ \citep{dal2021performance}} & \qquad Models \qquad\qquad& \qquad Number of parameters \qquad \qquad & \qquad RSEs\qquad\qquad \\
	\\
	\hline
	\\
	1 &Extended tube model& 3& 0.1921\\
	\\
	2 &Carroll model  & 3 & 0.2200 \\
	\\
	3&Xiang et al. model& 3 &0.2433\\
	\\
	4 &Davidson-Goulbourne model  & 3 & 0.3565 \\
	\\
	5 & \makecell{\textbf{Current model}\\$\rho k_BT=1.03$MPa; $N=149$} &\textbf{2}& \textbf{0.3736}\\
	\\
	\hline
\end{tabular}
\label{tab:rse}
\end{table*}

\section*{Appendix B}

For better clarification, the energy terms associated with corresponding Legendre transformations  is summarize as follows:\\
The Gibbs energy of a segment:
\begin{equation}
	\label{}
	G(f)=F-f\tilde b=k_BT\ln \frac{\xi}{\sinh\xi}.
\end{equation}
The Helmholtz energy of a segment:
\begin{equation}
	\label{}
	F(\tilde b)=G+f\tilde b=k_BT\left[\frac{\tilde b\beta}{b}+\ln\frac{\beta}{\sinh\beta}\right].
\end{equation}
The Halmiltonian (Gibbs energy formulated via new conjugate variable $(f\tilde b,\ln\lambda)$) of a segment:
\begin{equation}
	\label{}
	\tilde G=F-f\tilde b\ln\lambda=F-\dfrac{\partial F}{\partial \ln\lambda}\ln\lambda.
\end{equation}
The elastic energy density:
\begin{equation}
	\label{}
	W=\rho N\left<F\right>=\rho N\int_{|\vect u|=1} P(\vect u)\cdot F(\ln \lambda_u)\text d\vect u.
\end{equation}
The elastic complementary energy density:
\begin{equation}
	\label{}
\bar W=W-\vect\sigma:\vect h=\rho N\left<\tilde G\right>.
\end{equation}
The Helmholtz energy density of the network:
\begin{equation}
	\label{}
	\mathcal W=W-T\tilde S=W+k_BT\left<\ln P\right>
\end{equation}
The Gibbs energy density of the network:
\begin{equation}
	\label{}
\mathcal G=\mathcal W-\vect\sigma:\vect h=\bar W-T\tilde S
\end{equation}

\bibliography{ref1}

\end{document}